\documentclass[aps,prb,twocolumn, superscriptaddress,amsmath,amssymb, notitlepage,reprint,footinbib]{revtex4-1}
\usepackage[T1]{fontenc}
\usepackage[latin9]{inputenc}
\usepackage{amsfonts}
\usepackage{amsmath}
\usepackage{graphicx}
\usepackage{amssymb}
\usepackage{esint}
\usepackage{SIunits}
\usepackage{braket}
\usepackage{bm}
\usepackage{sidecap}
\usepackage{mathptmx}
\usepackage[breaklinks=true,colorlinks=true,linkcolor=red,urlcolor=blue,citecolor=blue]{hyperref}

\usepackage{color}

\newcommand{\rmnum}[1]{\romannumeral #1}
\newcommand{\Rmnum}[1]{\expandafter\@slowromancap\romannumeral #1@}

\begin{document}
\title{Tailoring topological states in silicene using different halogen-passivated Si(111) substrates}


\author{Vahid Derakhshan}
\email{riemann.derakhshan@gmail.com}
\affiliation {School of Physics, Damghan University, P.O. Box 36716-41167, Damghan, Iran}

\author{Ali G. Moghaddam}
\email{agorbanz@iasbs.ac.ir}
\affiliation{Department of Physics, Institute for Advanced Studies in Basic Sciences
(IASBS), Zanjan 45137-66731, Iran}
\affiliation{Research Center for Basic Sciences \& Modern Technologies (RBST), Institute for Advanced Studies in Basic Science (IASBS), Zanjan 45137-66731, Iran}

\author{Davide Ceresoli}
\email{davide.ceresoli@cnr.it}
\affiliation{CNR Institute of Molecular Science and Technology (ISTM), via Golgi 19, 20133 Milan, Italy}

\begin{abstract}
We investigate the band structure and topological phases of silicene embedded on halogenated Si(111) surface, by virtue of density functional theory calculations. 
Our results show that the Dirac character of low energy excitations in silicene is almost preserved in the presence of silicon substrate passivated by various halogens. Nevertheless, the combined effects of symmetry breaking due to both direct and van der Waals interactions between silicene and the substrate, charge transfer from suspended silicene into the substrate, and finally the hybridization which leads to the charge redistribution, result in a gap in the spectrum of the embedded silicene. We further take the spin-orbit interaction into account and obtain the resulting modification in the gap. The energy gaps with and without spin-orbit coupling, vary significantly when different halogen atoms are used for the passivation of the Si surface and for the case of iodine, they become on the order of $100$ meV. To examine the topological properties, we calculate the projected band structure of silicene from which the 
Berry curvature and $\mathbb{Z}_2$-invariant based on the evolution of Wannier charge centers are obtained. 
As a key finding, it is shown that silicene on halogenated Si substrates has a topological insulating state which can survive even at room temperature for the substrates with iodine and bromine at the surface. Therefore, these results suggest that we can have a reliable, stable and robust silicene-based two-dimensional topological insulator using the considered substrates.  
\end{abstract}

\maketitle

\section{INTRODUCTION}
Discovery of topological insulators (TIs) in last decade have introduced a new class of quantum matter
whose band structure has a nontrivial topological feature and subsequently it exhibit a metallic behavior in the boundaries while remaining insulating in the bulk \cite{Hasan-2010,Zhang-2011,bernevig-book}.
The topological protection in TIs is usually supported by some kind of symmetry like time-reversal (TR), parity, or number conservation. In most of the known TIs, the time reversal symmetry and spin-orbit interaction (SOI) play key roles in tailoring a material to its topological state as first pointed out by Haldane three decades ago \cite{Haldane}. In three dimension various materials including Bi$_2$Se$_3$, Bi$_2$Te$_3$, Sb$_2$Te$_3$, etc. have been proposed and eventually justified by experiments to be topological insulators. In the boundaries of these topological materials, two-dimensional (2D) metallic surface states establish which are governed by a single Dirac-cone spectrum \cite{Zhang-2009,Xia-2009}.
On the other hand, TIs in two-dimension possess robust helical edge states at the edges and usually show quantum spin Hall effect (QSHE) or other similar topological effects \cite{km-2005,wu-2006,qi-2010}. The most famous example of 2D TIs is HgTe quantum well (QW) in which the first reliable signature of QSHE has been reported almost a decade ago \cite{Bernevig-2006,Konig-2007}. Before HgTe QWs, graphene as the most well-known 2D material had been proposed to be a 2D TI which has helical edge states protected by TR symmetry \cite{kane-2005}. This fascinating proposal immediately boosted the search for 2D TIs, although it has been ruled out for the case of graphene itself due to the negligible SOI. In recent years, besides QW structures, various
atomically thin mono or few layers have been studied mostly theoretically to peruse nontrivial topological phases in 2D materials \cite{yao-prl-2011,Xu-2013,Ezawa-2012,Liu-2014,Chou-2014,tahir-2013,
fang-prx,nagaosa-2011,Wang-2013,zunger-2015,
Chuang-2013,Huang-2014,Ma-2014,Ding-2015,
Jin-2015,Kim-2016,Tan-2016,qian-2014,
cobden-2017,derakhshan}.
These atomically thin crystals usually consist of
either an element from group III to group VII or some certain transition metal compounds, although among them there are complex structures like organometallic lattices (for a decent review see Ref. \cite{niu-2016}). 
\par 
Among above-mentioned proposals for realizing topological phases in two-dimension, graphene-like materials which have slightly buckled honeycomb lattice and stronger intrinsic SOI are very promising \cite{yao-prl-2011,Xu-2013}. In particular silicene, germanene, and stanene which are graphene successors made of Si, Ge, and Sn atoms respectively, are believed to undergo a transition to topologically nontrivial states at least in some engineered forms \cite{Liu-2011, huang-jmc-2012}. In particular, silicene despite its small SOI ($\sim 1$ meV), has the great advantage of being compatible with the existing silicon-based electronic industry. 
In contrast to graphene, free-standing silicene (FSS) exhibits a Dirac electron structure around the two $K$-points is very elusive \citep{Vogt}.
So far, silicene was synthesized on various supporting substrates including Ag$(111)$ \cite{lalmi2010}, ZrB$_2$(0001) \cite{Ozaki}, and Ir(111) \cite{Meng}. Unfortunately, using metallic substrates, significantly changes or even destroys the 2D Dirac dispersion of silicene due to the hybridization with electronic states of metal at the vicinity of Fermi level \cite{Mahatha}. Furthermore, such substrates fully screen the externally applied electric fields which prohibit the engineering of electronic states and also the application in field effect transistors. Therefore, quasi-FSS grown over insulating materials where van der Waals (vdW) interaction acts between silicene and the substrate,
is highly demanded. Besides the insulating nature, any eligible substrate for growing silicene which can hopefully make the observation of QSHE possible, should have hexagonal structure at the surface commensurate the lattice of silicene with a tiny mismatch. While no insulating substrate has been experimentally found yet, various theoretical works have
examined possible insulating substrate including hexagonal monolayer boron nitride, SiC, and silicon by virtue of first principles calculations\cite{Liu-1, Schwingenschlogl,Kaloni,Kokott-1,
Kokott-2,Wang,Liu-2,gao-2012}.
\begin{figure}[t]
	\centering
	\includegraphics[width=0.95\linewidth]{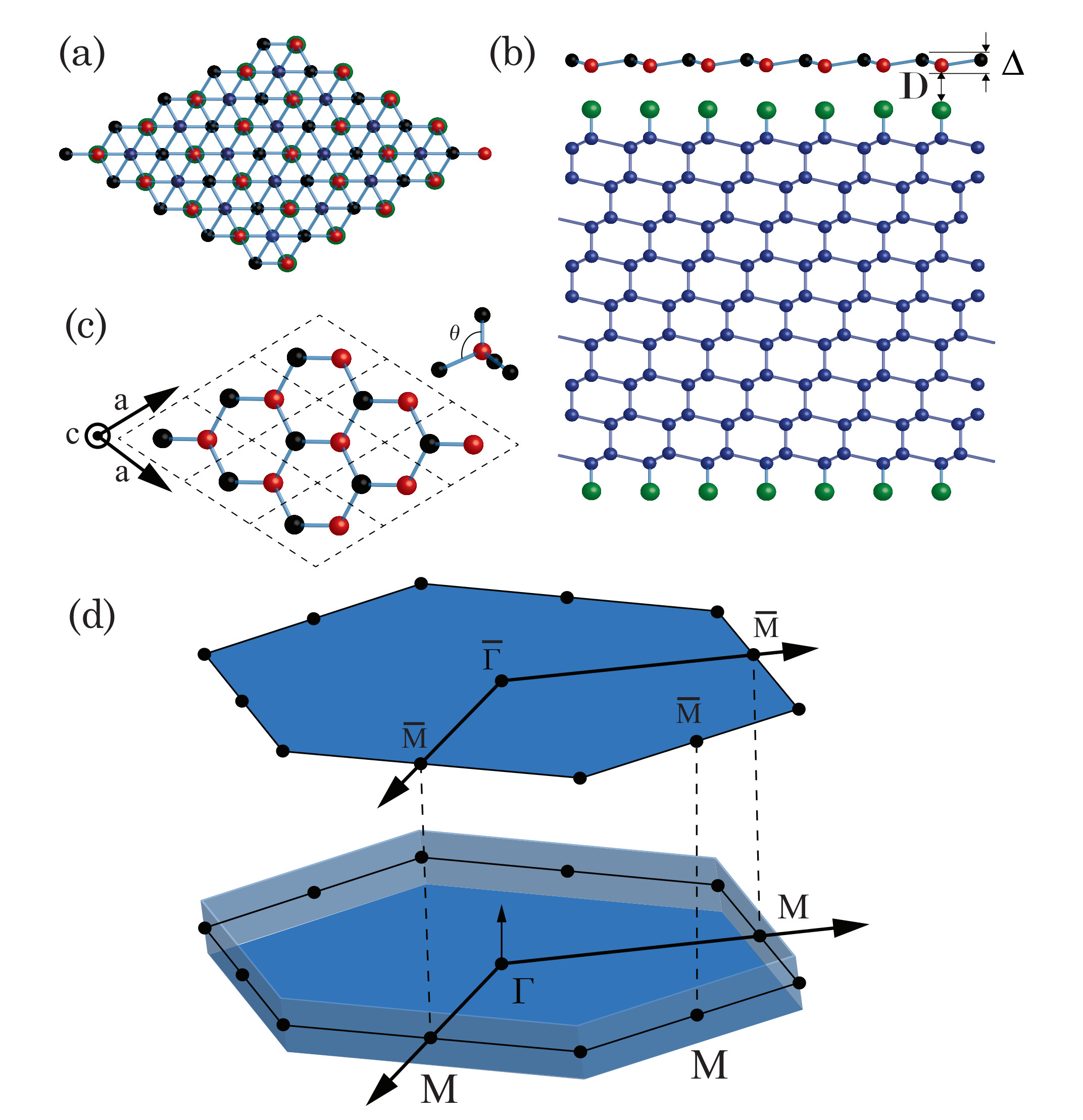}
	\caption{\label{fig1} (Color online)
Schematic view of silicene on top of the halogenated Si($111$) surface: (a),(b) show top and side views, respectively. Blue spheres indicate silicon atoms from Si($111$), red and black balls show silicon atoms from silicene's two sublattices, respectively. Green balls indicate the halogen atoms used for passivation of the silicon layer. 
(c) Free-standing silicene with two silicon atoms in unit cell. (d)  Brillouin zone and its 2D projection for the hexagonal crystal structure in which high symmetry points are indicated.}
\end{figure}
\par
In this work, we will investigate the electronic band structure of silicene supported on halogenated Si$(111)$ surface in the framework of density functional theory (DFT). Our results show that Dirac nature of silicene is not influenced significantly by the presence of the Si substrate with absorbed F, Cl, Br or I atoms at the surface. Nevertheless, 
due to the difference in the on-site energies of A and B sublatices of suspended silicene induced by direct interactions and mediated by weaker can der Waals 
forces, a small charge transfer from silicene into the substrate, and also hybridization of electronic clouds between silicene and the substrate, a gap ($E_{g1}$) of $12-68$ meV depending on the type of halogen atoms is opened up at the Dirac point of silicene. By turning on the SOI, the gap can be varied by an amount varying between $-10$ to $14$ meV
for different halogen atoms used for the passivation.
In addition to the total band structure of the silicene/substrate composite structure, we further calculate the atomically projected band structure for the silicene monolayer itself. Then we use the atomically projected energy dispersion to obtain hopping and overlap parameters of the effective  Hamiltonian for silicene on the substrate. To examine the topological properties of the embedded silicene layers, we calculate the Berry curvature and alternatively the evolution of so-called \emph{Wannier charge centers} (WCCs) and the largest gap in their spectrum
\cite{Alexey}. 
These methods have the advantage of obtaining the topological aspects directly from the ab initio results and the Wannier interpolation of the band structure obtained from \emph{WannierTools} post processing package \cite{WannierTools}. 
Both the Berry curvature calculations and WCCs evolution, reveals that the embedded silicene over the halogenated silicon layer preserves its nontrivial topological nature the same as FSS. On the other hand, the stability of the proposed structures besides the larger gaps induced in the embedded silicene 
suggest that the silicene on hologenated Si substrate can be considered as a stable and more robust 2D topological insulator. 
\section{ Model and computation framework}
\label{sec-2}
We consider a quasi-FSS monolayer over a silicon substrate which is passivated by various halogen atoms at Si(111) surfaces (see Fig. \ref{fig1}). 
In order to implement DFT calculations we assume that the Si layer consists of 9 slabs which leads to the thickness of $\sim3$ nm for the substrate. In order to mimic a semi-infinite substrate which is closer to the realistic situations with many slabs, we use constrained relaxation by fixing the three bottom slabs of Si(111).
As we have already mentioned the passivation of silicon surfaces with halogens is required to prevent the formation of direct covalent bonds (with length $\sim2.3$ \AA) between atoms of silicene and the substrate. Subsequently, silicene will be suspended over the substrate by a larger distance of $D\gtrsim3$ \AA~ with weak vdW interaction between them. 
\par
The \emph{ab initio} calculations are done using
Quantum-ESPRESSO package \cite{qe}, mainly in the framework of generalized gradient approximation (GGA). The vdW interaction is taken into account via vdW-DF-ob86 functional to implement vdW DFT calculations \cite{Thonhauser-1,Thonhauser-2}. 
In our calculations, in order to have comparison, the band structure will be obtained both at the presence and absence of SOI. The plane-wave cutoff energy is considered to be $540$ eV which is obtained via optimization. For the self-consistency calculations, a Monkhorst-Pack of $21\times21\times1$ for the k-mesh is employed. By full relaxation of the lattice structure, we find in-plane lattice constant to be $\textbf{a} = 3.90$ \AA. Moreover, a vacuum space
of $40$ \AA~ is set to prevent the interaction between the periodic images of the whole substrate/silicene layer along the $c$-axis. For the whole ab initio calculations, an energy convergence of $10^{-8}$ eV and a force convergence of $0.01$ eV/\AA~ are achieved.

\begin{table*}[htb]
 	\centering
 	\caption{Lattice structure and band dispersion parameters of silicene supported on the halogenated substrate besides corresponding values in FSS for comparison. Lattice constant a, buckling amplitude $\Delta$, formation energy $E_{form}$ (eV/Si atom), Si-Si bond length and the distance D between the bottom silicene atoms with the halogen layer are obtained from total energy DFT calculations including vdW interaction. $E_{g1/HSE06}$ was calculated with HSE06 functionals for comparison The other parameters are related to the band structure of silicene near the $K$ point. The units of Fermi velocity $v_F$, energy gaps $E_{g1,2/PBE}$ and $E_{Dirac}$ are $10^5$ m/s, meV respectively.}
 	\label{tab:structureparameter}
 	\begin{tabular}{cccccccccccccc}
 		\hline  \hline
 		Atom & a (\AA) & $\Delta$(\AA)& $E_{form}$ & Si-Si(\AA)& D(\AA)& $\theta$(deg) &$\nu_{F}$(TB) & $\nu_{F}$(DFT) & $E_{g1/PBE}$&$E_{g1/HSE06}$& $E_{g2/PBE}$ & $E_{Dirac}$\\
 		\hline
 		--      & ${3.86}^a$    & ${0.46}^a$  & --     & 2.280    & --     & ${101.7}^a$  & ${5.52}^a$ & ${5.42}^a$ & 0    & 0  &   ${1.50}^a$   & --      \\
 		F       & 3.90          & 0.42        & -2.85  & 2.295    & 3.11   &  100.6       & 4.71       &  5.35      & 40   & 77 &   30.0         & +29   \\
 		Cl      & 3.90          & 0.42        & -3.18  & 2.296    & 3.60   &  100.7       & 5.04       &  5.20      & 12   & 43 &   20.0         & +17   \\
 		Br      & 3.90          & 0.43        & -3.21  & 2.297    & 3.47   &  100.8       & 5.05       &  5.10      & 47   & 104 &   59.0         & +1   \\       
 		I       & 3.90          & 0.47        & -3.19  & 2.304    & 3.31   &  101.7       & 5.07       &  4.90      & 68   & 139 &   82.0         & -2   \\
 		\hline\hline
 	\end{tabular} \\
 	a: for free-standing silicene {Reference~\cite{Liu-2011}.}
 \end{table*}
\par
In order to concentrate only on the electronic properties of the silicene and subsequently obtain the topological aspects in the band structure we use post-processing WANNIER90 package \cite{wannier90-1,wannier90-2}, to construct the maximally localized Wannier functions (MLWFs) which are centerred at Si atoms of silicene. Then based on the Wannier-interpolated band dispersions, we proceed to explore how Si($111$) substrate passivated by various halogen atoms, impacts the band topology of suspended silicene layer. Topological properties of the silicene valence bands can be investigated based on the distribution of Berry curvature  corresponding to these energy bands over the whole Brillouin zone. This is achieved by by utilizing WannierTools post processing package \cite{WannierTools}.
The non-trivial behavior of topological insulating band structure can be usually characterized by the so-called  $\mathbb{Z}_2$ topological invariants. At the presence of TR symmetry the 2D spinful system like silicene belongs to the class AII of topological phases~\cite{Schnyder08,Ryu10} and depending on the value of $\mathbb{Z}_2 =0,1$ it has a trivial and non-trial topological state, respectively. When the inversion symmetry is present, the topological invariant can be directly obtained from the
product of parity eigenvalues for all the occupied bands at the time-reversal-invariant momenta (TRIM) points, which has been widely used for 2D topological insulators \cite{kane-2007,Zhang-2013,
Zhang-2015,Bansil-2014,Cahangirov-2014}.
On the other hand for the systems in which the inversion symmetry is broken, more complicated analysis is needed for the calculation of $\mathbb{Z}_2$. In these cases, following the Wannier function approach proposed by Soluyanov and Vanderbilt, the topological aspects can be obtained by tracking the evolution of  WCCs and their largest gap \cite{Alexey}.
\section{Results and discussions}
\subsection{DFT calculations and band structure}
We start by optimization of the primitive cell of silicene supported on halogenated Si($111$) surface. As a result, a lattice parameter of $\textbf{a} = 3.90$ \AA~ and Si-Si bond length $2.28-2.30$ \AA~  depending on halogen atoms are obtained, which have very good agreement with the previous theoretical investigations \cite{Cahangirov,Houssa,Eriksson}. As it can be seen from Fig. \ref{fig1}(b), both sides of Si($111$) surface are passivated with the halogen atom (green balls). The halogenated substrate is perfectly lattice-matched to the Bravais lattice of the silicene sheet with a deviation of only $0.1\%$. After relaxation, the silicene sheets stablizes on the substrates with an equilibrium distance $(D)$ depending on the type of halogen atom. We find that buckled height ($\Delta$) of silicene layer supported on the substrate varies slightly for different halogen atoms. When fluorine, chlorine, bromine and iodine atoms are used for passivation, the bucklings $0.42$, $0.42$, $0.43$ and $0.47$ \AA~ are obtained, respectively, different from the natural buckling $0.46$ \AA~ in the case of FSS. Nevertheless, these results reveal that the substrate can provide an effective mechanical support without significantly disturbing of the lattice parameters of silicene. The structural information for silicene on the halogenated substrates are listed in Table~\ref{tab:structureparameter}.
\par 
In order to investigate the stability of the obtained structure after relaxing it, we calculate the formation energy of silicene/substrate composite system which is defined as \cite{silicenemos2},
\begin{equation}\label{formation}
E_{form} = (E_{\rm Si/sub} - E_{si} - E_{\rm sub})/N_{\rm Si}
\end{equation}
where $E_{\rm Si/sub}$, $E_{\rm Si} $ and $E_{\rm sub} $ represent the total energy of the silicene/substrate composite, isolated silicene, and the isolated substrate, respectively. $N_{\rm Si}$ is the number of Si atoms in the isolated silcene layer. The calculations show that the formation energy of considered system per silicon atom  when F, Cl, Br and I atoms are used for passivation are $-2.85$, $-3.18$, $-3.21$ and $-3.19$ eV which are one order of magnitude larger than similar systems, such as recently proposed silicene/MoS$_2$ composite structure \cite{silicenemos2}. Indeed, this means that the silicene/substrate structures considered here are energetically more stable than similar ones which use  different substrates. As can be seen from Table \ref{tab:structureparameter}, except for Florine-passivated case, formation energy for other halogen passivated case are nearly the same while the Bromine passivated case has the largest formation energy. 
\begin{figure*}[!htb]
	\centering
	\includegraphics[width=0.9\textwidth, angle = 0]{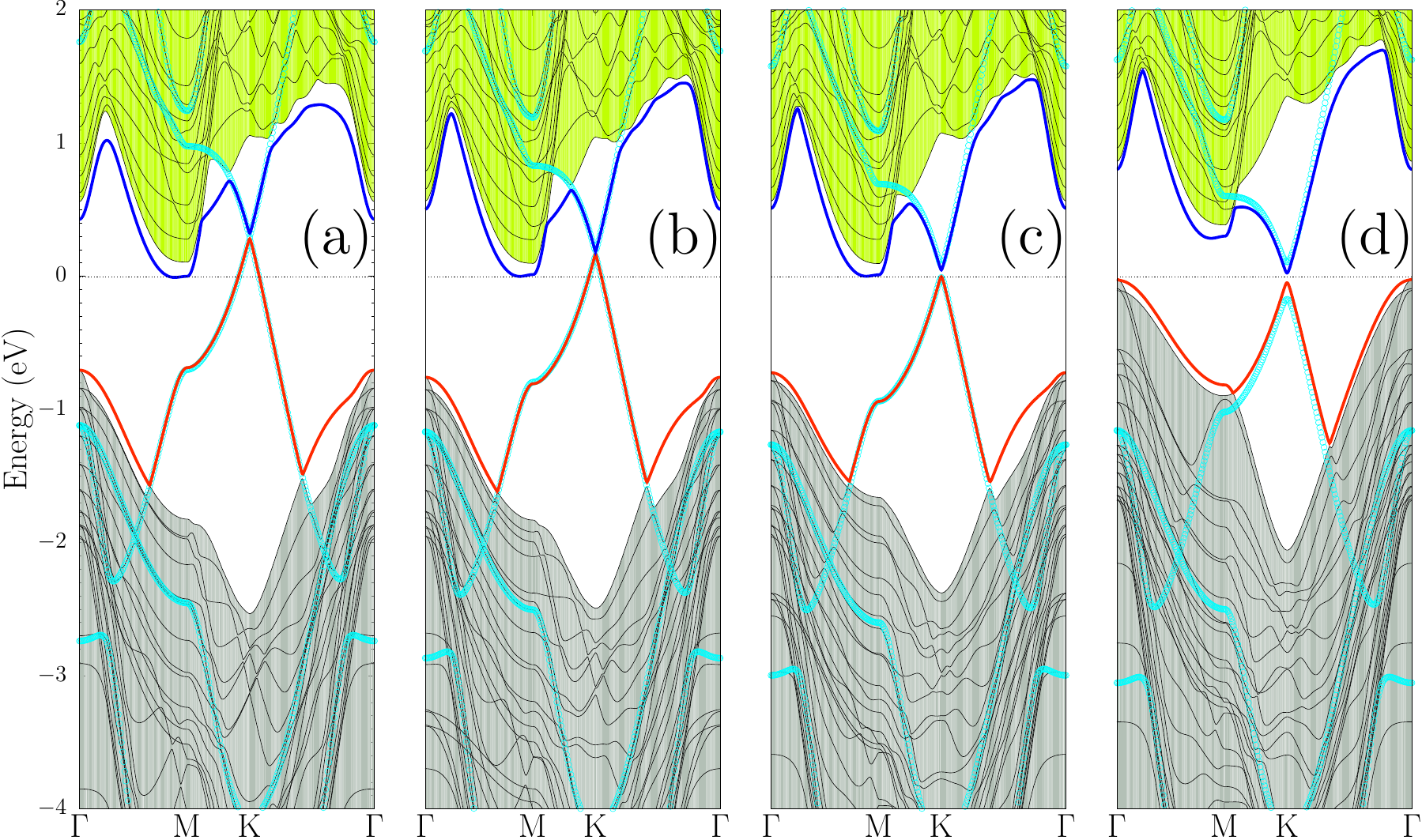}
	\caption{\label{fig2} (Color online) Calculated band structure of silicene supported on the halogenated Si($111$) surface. Solid black lines: original bands generated directly from DFT calculation. Solid red and blue lines indicate valance and conduction bands, respectively. Cyan circles: Wannier-interpolated bands obtained from the subspace selected by an unconstrained projection onto atomic $sp^3$ orbitals of silicene. (a)-(d) panels corresponding to fluorine, chlorine, bromine and iodine passivated cases, respectively.}
\end{figure*}
\par 
Figure~\ref{fig2} presents the  calculated band structure of silicene/halogen passivated  Si($111$) substrate. The solid black lines indicate the band structure of silicene/substrate system, obtained from DFT and cyan circles show the band structure of only the silicene layer reproduced using the WANNIER90 package
using a $36\times36\times1$ k-space sampling density.  The band structures shown in  Fig. \ref{fig2} confirms the claim that the Dirac cones are preserved when silicene is supported on the halogenated Si($111$) substrate. Careful examination of the detailed band structure at the $K$-points of the Brillouin zone reveals that a tiny gap ($E_{g1}$) is opened 
with different values depending on the passivating halogen as presented in Table~\ref{tab:structureparameter}. Moreover, to justify the emergence of the gap due to the coupling of silicene and the substrate, the calculations may be additionally performed in the framework of HSE06 hybrid functional. To this end, we use CRYSTAL14 package \cite{CRYSTAL14} with the same k-point mesh and with the optimized all-electron gaussian basis sets of Mike Towler~\cite{towler-basis} and the resulting values for the gap $E_{g1/HSE06}$ are shown in Table~\ref{tab:structureparameter}. We see that Crystal14 calculations result in larger gaps by amount $20-40$ meV in comparison with GGA-based method. The appearance of the gap $E_{g1}$ compared to the pristine silicene,
can be related to various reasons  explained as what follows: 

(\rmnum{1}) While in case of FSS, the gap is induced only by effective spin-orbit interaction, here, it can partially originates from the broken inversion symmetry between different sublattices of silicene. In fact direct interaction between the electronic clouds of the substrate and silicene at the interface which can be accompanied by weaker vdW interactions, results in an effective staggered potential acting on the silicene electrons in different sites.
Such a potential and corresponding symmetry breaking leads to a gap between touching conduction and valence bands of silicene. 

(\rmnum{2}) Charge redistribution over silicene layer which originates from hybridization between the closest orbitals from silicene and the halogenated substrate electronic states. 

(\rmnum{3}) A tiny charge transfer at the interface in accompany with other effects mentioned above builds up the interlayer potential difference between the two sublattices of silicene which can enhance the gap \cite{Zhou-2014}.
\begin{figure}[!htb]
	\centering
\includegraphics[width=0.95\columnwidth, angle = 0]{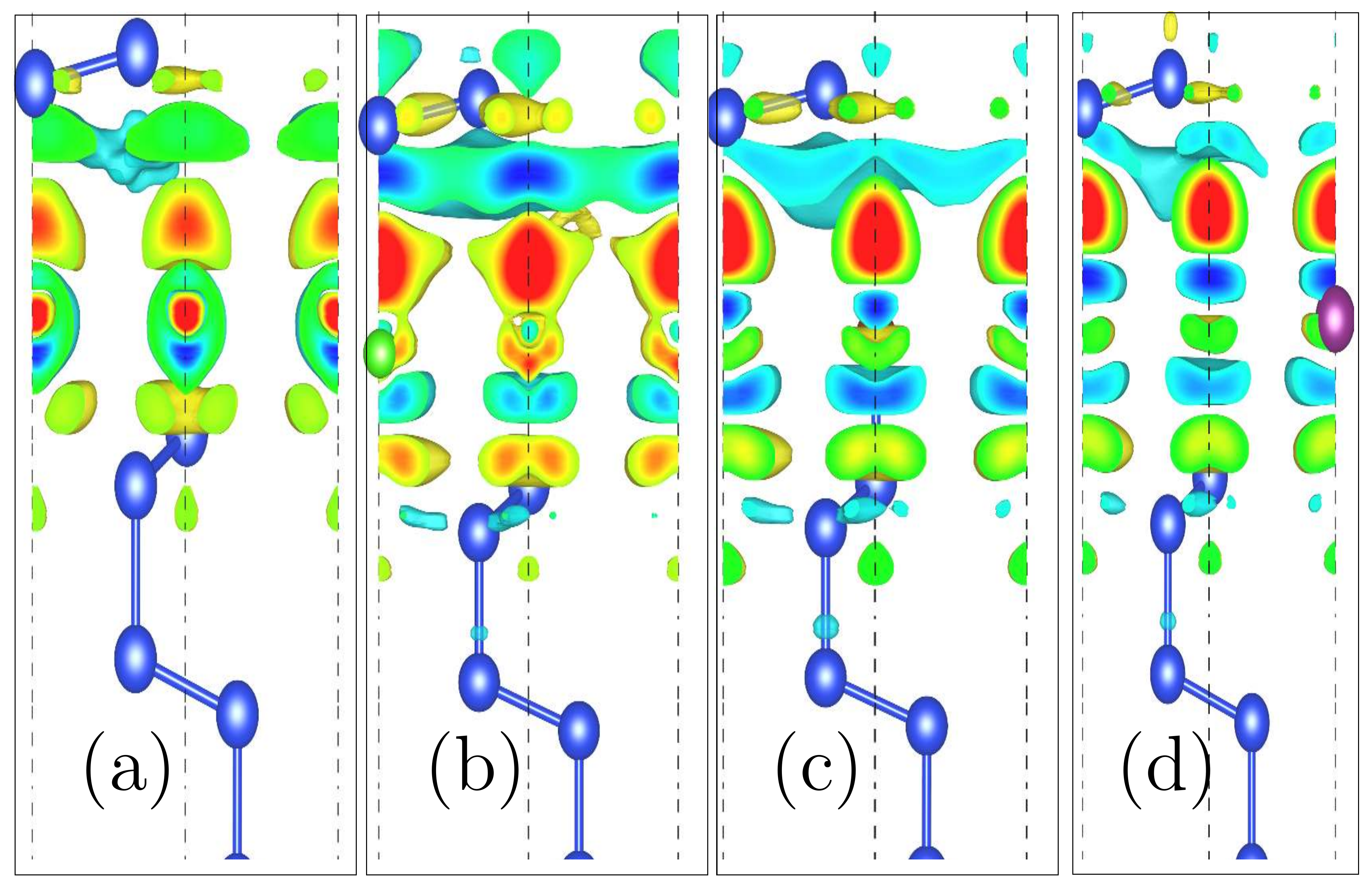}
	\caption{\label{CDD} (Color online) Charge density difference of silicene on halogen passivated Si (111) surface. Panels (a-d) corresponding to the cases when F, Cl, Br and I atoms are used for passivation, respectively.}
\end{figure} 

\begin{figure*}[t]
\centering
\includegraphics[width=0.9\textwidth, angle = 0]{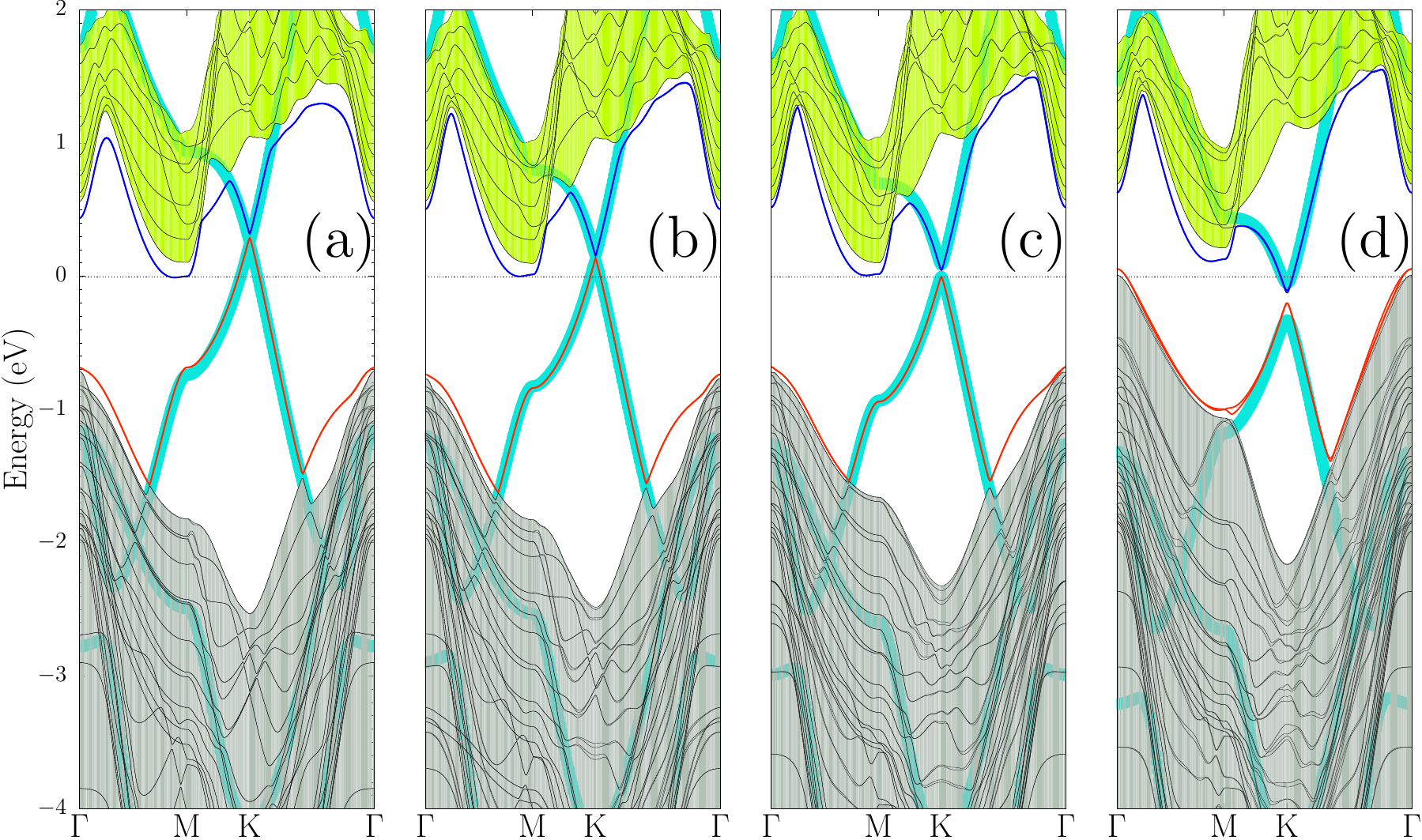}
\caption{\label{fig4} (Color online) Band structure of silicene  supported on halogenated Si($111$) surface in the presence of SOI.(a)-(d) correspond to fluorine, chlorine, bromine and iodine passivated substrates. Cyan circles: Wannier-interpolated bands obtained from the subspace selected by an unconstrained projection onto atomic $sp^3$ orbitals of silicene.} 
\end{figure*}
\par
It worth to note that while the vdW interaction increases the adsorbtion energy and bring silicene closer to the substrate but it does not play an important role in gap opening compared to the charge redistribution and direct interaction between the silicen/substrate electronic clouds. To justify this we use GGA and LDA pseudo-potentials without vdW corrections in fixed separation distance between silicene and substrate. The results show that the band gap remains nearly the same to the case in which vdW corrections pseudo-potentials have been used. The charge redistribution in interlayer interval and silicene layer is another source incorporating in the gap opening at the vicinity Dirac cones. This can be seen by  investigating the interlayer interaction via the charge density difference (CDD) $\Delta\rho = \rho_{\rm Si/sub} - \rho_{\rm sub} - \rho_{\rm Si}$ between the situation with substrate and the silicene layer close to  and isolated from each other.
The results of these calculations in which the pseudo-potentials without vdW corrections are used, can be seen from Fig. \ref{CDD}. It should be mentioned that in order to have a correct comparison, we invoke the same parameters for the isolated silicene and substrate system as the hybrid one. 
The CCD plots reveal that the redistributed charge mainly accumulates in the interval between the silicene layer and the Si substrate for all kind of halogens used for the passivation. 
\par
As we have mentioned above the charge redistribution mostly originates from the direct hybridization between the electronic orbitals of Si atoms in silicene and those in the substrate. Combined with the interaction effects this leads to the staggered potential and subsequently a gap opens. Therefore we can conclude that as a rule of thumb the larger charge imbalance between the two sublattices would give rise to a larger gap at the Dirac point of Silicene's band structure. On the other hand, Table~\ref{tab:structureparameter} shows that the gap $E_{g1}$ monotonically increases
by using heavier halogen atom for the passivation of the Si layer, if we ignore the transition from F- to Cl-passivated cases. This can be explained, indeed, based on the origins of gap opening which include
the interlayer coupling or interactions, charge redistribution and tiny charge transfer between silicene and the substrate. In fact, in case of Florine its huge reactivity plays a major role in interlayer interaction rather than vdw interaction due to its small atomic radius. But, in three other cases, considering that with increasing the atomic radius, their reactivity reduces while vdw interaction gets slightly enhanced which brings the suspended silicene closer to the substrate.
Therefore the hybridization between the orbitals of silicene and those at the surface of the substrate and the resulting symmetry breaking effects become more profound and a larger gap appears in the band structure.  
\par
Now we turn to the Dirac picture of the low energy bands of the silicene/substrate hybrid system. As introduced in Ref. \onlinecite{Liu-2011}, the Fermi velocity in 2D buckled structures can be obtained by a relation based on fitting Slater-Koster parameters \cite{slater}. The estimated Fermi velocity of suspended silicene from DFT results 
using $ \nu_F \approx (1/\hbar) (dE/dk)$ and also TB methods, can be seen in Table~\ref{tab:structureparameter} which again show good agreement with previous investigations \cite{Liu-2011}. The invariability of Fermi velocity
further illustrates that the halogenated/Si$(111)$ surface does not profoundly disturb the electronic and transport properties of suspended silicene layer. This indeed justifies the fact that halogenated Si(111) can be considered as a promising noninvasive substrate for silicene. As is shown in panels (a-d) in Fig. \ref{fig2} for silicene/Si$(111)$ systems, the characteristic electronic properties of silicene which resulted from the
linear dispersion around Fermi level and the existence of
the Dirac cone, are almost preserved. In other words, electronic band structure of silicene behaves almost the same for all of the halogen atoms with some trends from fluorine to iodine. It is well known that the pristine silicene is a semimetal, where the $p_z$ and $p_z^*$ orbitals give rise to $\pi$ and $\pi^*$ bands and as a result at the $K$ and $K'$ points and exactly in the Fermi energy 
two Dirac cones are formed \cite{Ezawa-2012,Vogt}. For the silicone substrates in which fluorine, chlorine and bromine atoms are used for the passivation, due to the positive energy shift of the Dirac cone, the conduction band of silicene is partially mixed with upper energy bands of the substrate [see Figs. \ref{fig2}(a)-(c)]. The energy shifts of Dirac point with respect to Fermi energy denoted by $E_{\rm Dirac}$, are listed in  Table~\ref{tab:structureparameter}
for all of halogen atoms.
Very intriguingly when iodine atoms are positioned over the Si surface, the Dirac cone of silicene layer is shifted down and subsequently it lies within the band gap of the substrate, as can be seen from Fig. \ref{fig2}(d).
\par
As it has been discussed in the introduction besides the low-energy Dirac spectrum, in fact, the main ingredient of the topologically nontrivial band structure is SOI which leads to the QSHE in two dimension. 
Therefore before starting to investigate the topological aspects we should include the SOI effects in the band structure. 
Figure \ref{fig4} shows the band structures of silicene/substrate system plus atomic-projected (denoted by cyan circles) band structure of suspended silicene in the presence of SOI.
We should remind that for the sake of comparison we calculate the band structure both at the presence at absence of SOI which are shown in Figs. \ref{fig4} and \ref{fig2}, respectively. As expected, explicit inclusion of SOI in the DFT calculations lead to a gap $E_{g2}$ which is typically a few meV larger than $E_{g1}$ depending on the halogen atoms  as listed in Table \ref{tab:structureparameter}. We should note that the case of fluorine is exceptional in this regard and inclusion of SOI has reduced the gap. 
Similar to our previous results, even at the presence of SOI, while usage of fluorine, chlorine, and bromine atoms for the passivation leads to the hole-doped silicene, iodine exceptionally results in  
a weak electron doping due to the negative energy shift of the Dirac cone (see Fig. \ref{fig4}). Moreover, from Figs. \ref{fig2}(a)-(c) and \ref{fig4}(a)-(c), one can see that the conduction band minimum (solid blue line) belonging to the substrate bands around the M-point is shifted downward and crosses the Fermi level. Therefore we can conclude that the positive energy shift of the Dirac cone and the negative shift of the substrate bands close to the M-point are consistent with the charge transfer between the silicene and the substrate in a consistent manner.  
In particular, the hole-doping of silicene is enhanced for fluorine and chlorine cases, along with the 
further downward energy shift
in the conduction band minimum within the M-point.

Although the Dirac point in the hybrid silicene/substrate structures considered here is displaced above or below the Fermi level (positive and negative $E_{Dirac}$) depending on the substrate passivation, however it can be easily replaced by small amount of chemical doping or by applying external gate voltages.
In fact the 2D and layered character of the system make it very tunable and by both chemical and electric methods it can be tailored much easier than bulk structures to reach a desired electronic properties. So in principle one can externally tune the position of Dirac point with respect to the Fermi energy to lie deep inside the energy gap of the silicene/substrate complex structure.
There is no need to emphasize that this tunability is very crucial to engineer a reliable topological insulating phase.
In fact when the states from the substrate lie within the Fermi level even if they do not have significant effect on the states at K-point and their nontrivial topology is preserved, it will be very unlikely to recognize topological phase of embedded silicene by transport measurements. Therefore in order to have a feasible true topological insulating states besides having a nontriviality at the electronic states of embedded silicene (at the vicinity of Dirac point) we should get rid of other states coming from the substrate itself at the Fermi energy which is possible by external gates or chemical doping in our system.

\begin{figure}[t]
	\centering
	\includegraphics[width=0.98\linewidth, angle = 0]{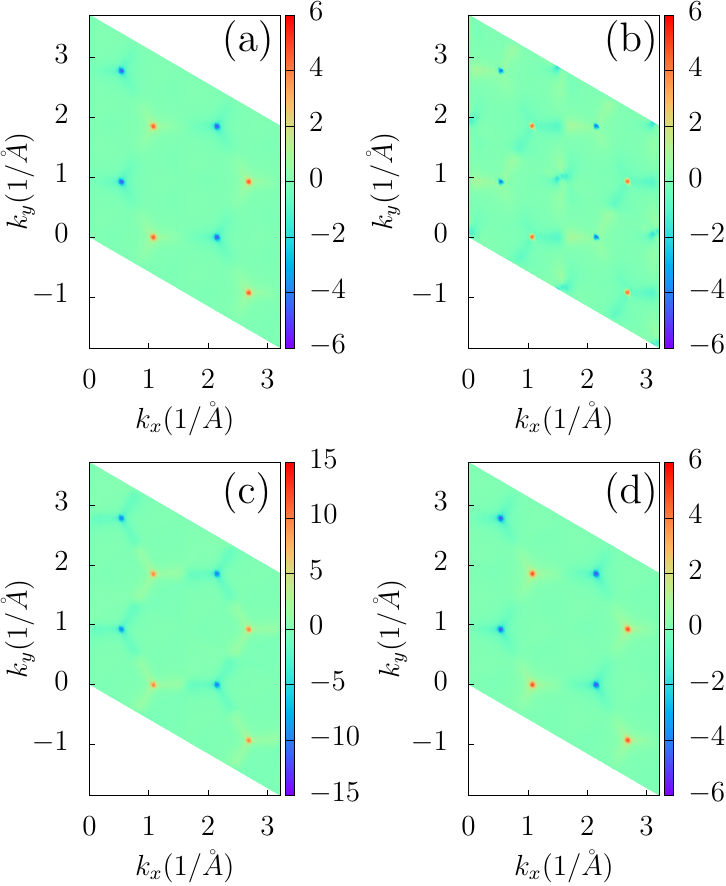}
	\caption{\label{fig3} (Color online) Calculated Berry curvature distribution of valence bands  below Fermi energy in $({k}_x,{k}_y)$ plane in arbitrary unit. (a)-(d) correspond to the fluorine, chlorine, bromine and iodine passivated substrates. These plots clearly show that the TR symmetry is not destroyed by the presence of the substrate and corresponding interaction and overlap effects with silicene atoms}
\end{figure}

\subsection{Topological properties}
\begin{figure*}[htp]
	\centering
	\includegraphics[width=0.9\textwidth, angle = 0]{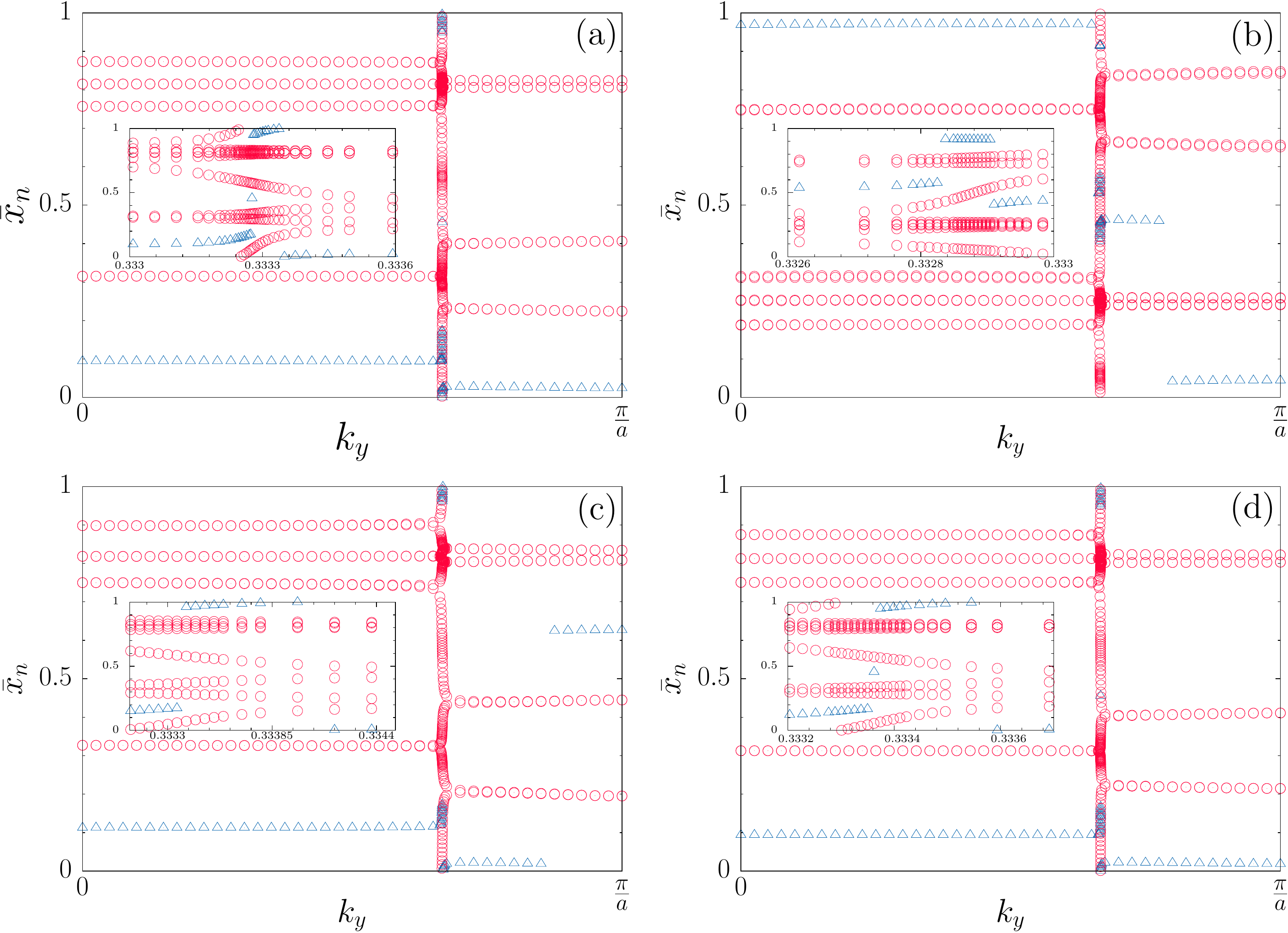}
	\caption{\label{fig4-1} (Color online) Evolution of WCCs ($\bar{x}_n$) for silicene at ${k}_z = 0$ plane as a function of ${k}_y$. Red circles indicate the WCCs calculated by WannierTools package. Blue triangles marks midpoint of largest gap. (a-d) panels are corresponding to the fluorine, chlorine, bromine and iodine passivated cases, respectively. Due to the small gap of the system there is a region of dense points and very sharp variations in WCCs which can be seen with better resolution inside the insets of each plot. 
It can be seen that in all cases the path of largest gap middle (blue triangles) crosses an odd number of WCC bands which identifies a topologically nontrivial phase protected by TR symmetry.} 
\end{figure*}

We proceed to explore how Si($111$) substrate passivated by various halogen atoms, impacts the band topology of suspended silicene layer. Topological properties of the silicene valence bands can be partially investigated based on the distribution of Berry curvature  corresponding to these energy bands over the whole Brillouin zone \cite{haldane-prl06,sheng-prl11,ezawa-prl12,Vanderbilt}. By utilizing WannierTools post processing package, we calculated Berry curvature for the band structure corresponding to the Hamiltonian based on MLWFs which are provided by WANNIER90 package. As indicated in Fig.~\ref{fig3}, the Berry curvature of suspended silicene band structure behaves almost the same as FSS and does not crucially change with the type of halogen atoms used for the passivation. One can see that while Berry curvature vanishes inside most of the Brillouin zone, it has very strong peaks with opposite signs around the Dirac points 
$K$ and $K'$. The existence of this singular peaks around $K$ and $K'$ in the Berry curvature shows the survival of TR symmetry of the silicene over the substrate. This itself suggest that not only the band structure but even
the topological properties of the embedded silicene is most probably  preserved as FSS, since the topological phase of silicene which belongs to class AII, is protected by the presence of TR symmetry. 
\par
In order to provide smoking gun prove for the nontrivial topology of the low energy bands in silicene/substrate hybrid system we should identify the $\mathbb{Z}_2$-invariant as we have discussed before in Sec. \ref{sec-2}. So we implement the calculation for WCCs to track their largest gap, following the method of Ref. \onlinecite{Alexey}. This method 
exploits the concept of TR polarization previously introduced by the analogies of a 2D TR-invariant topological insulator and a TR-symmetric pumping process \cite{kane-prl06}. Then the idea of TR polarization is reformulate in terms of the winding of the WCCs around the BZ.
Very intriguingly it has been shown that the $\mathbb{Z}_2$-invariant can be computed only by tracking midpoint of the largest gap between any two WCCs. In fact, the topological invariant is nothing but $\mathbb{Z}_2={\mathbb J}~mod~2$ with the integer number $\mathbb{J}$ indicating how many times the path of the largest gap's midpoint jumps over WCC bands which are normalized to $[0,1)$. Therefore the system is in a topological state as long as WCC bands are crossed an odd number times by the path corresponding to the middle of their largest gap. Since the considered silicene/substrate system is not periodic in $z$ direction. Thus, it is enough to compute only one $\mathbb{Z}_2$ index, say at ${k}_z = 0$. On the other hand, if we assume the momentum $k_y$ as the pumping parameter, it is sufficient to calculate WCCs only over the half of the range of momenta in Brillouin Zone
due the TR symmetry and the fact that any $k_y$ point has a TR partner at $-k_y$.
So the computational package WannierTools is able to calculate the topological $\mathbb{Z}_2$-invariant by utilizing the MLWF-based Hamiltonian obtained from  WANNIER90. For the hybrid silicene/Si(111)-substrate system considered here with different passivations of the substrate by 
F, Cl, Br and I atoms, the evolution of WCCs and the middle path of the largest gap are shown in Figs. \ref{fig4-1}(a)-(d), respectively. Our calculations result in $\mathbb{Z}_2 = 1$ as a direct output of WannierTools package which show that for all cases with different halogens used for the passivation, the nontrivial topology in the silicene's band structure is preserved. This result can also be justified by careful inspection of the evolution of WCC bands in Fig. \ref{fig4-1} even by bare eye instead of the computer code. 
\par
At the end, we should comment on the experimental feasibility of the proposed hybrid structure of silicene on halogenated Si(111) substrate and the predicted topological insulating phase. As it has been mentioned in the introduction, many theoretical and experimental efforts focused on putting silicene over the noninvasive substrate where the low energy features of silicene are not significantly affected. Although there are some successful experiments including those with Ag or MoS$_2$ substrates, the search for alternative substrate is still very active and particularly those which keep the Dirac dispersion as well as topological nontrivial phase are highly demanded. So our proposed systems which shows a very good stability according to their formation energies and predicts the existence of topological phases in this kind of silicene/substrate hybrids should be interesting for the experimental investigation. We should remind that our study reveals that the presence of hologene atoms which decreases strong overlap and hybridizations between the substrate and silicene plays an important role in preserving the topological properties of the embedded silicene layer. In addition, we find that the weak interactions of direct or vdW type as well as hybridization and charge redistributions are sufficient enough to open a gap of $\lesssim 100$ meV at the vicinity of Dirac point which enhances the feasibility of the topological signatures particularly in transport experiments.  

\section{Conclusions}
To summarize, we have investigated the electronic band structure and the  topological aspects in the hybrid system composed of a silicene layer supported on halogenated Si$(111)$ surfaces using the Density functional theory. To this end, we have performed the ab initio calculations for the band structures of suspended silicene over differently passivated substrates. Then using the WANNIER90 packages the atomic projected band structure of silicene have been obtained from which the topological properties can be followed. Our results show that the Dirac nature of silicene is not influenced significantly by the halogenated Si substrate. However, due to the direct interactions and hybridizations as well as van der Waals forces which makes the two layers closer, we see charge redistribution between different atomic sites as well as effective staggered potential which leads to a gap of $10-60$ meV around Dirac point as expected from symmetry breaking. Inclusion of spin-orbit interaction leads to a small variation of the gap with a range on the order of $\sim 10$ meV. 
Investigations based on the Berry curvature, $\mathbb{Z}_2$-invariant calculations based on the evolution of Wannier charge centers, reveal that not only the substrate does not destroy the nontrivial topological phase which has been predicted earlier for free standing silicene, but the robustness of topological state is enhanced. 

\begin{acknowledgments}
V.D. would like to thank M. Ghaebi and QuanSheng Wu for helpful discussions. 
A.G.M. acknowledges financial support from Iran Science Elites Federation under Grant No. 11/66332.
We acknowledge the CINECA award Nr. HP10C7BPGD under the ISCRA initiative,
for the availability of high performance computing resources and support.
\end{acknowledgments}


\end{document}